\documentclass[a4paper,12pt]{article}
\usepackage[utf8]{inputenc}
\usepackage{natbib}
\usepackage{graphicx}
\usepackage{longtable}

\title{Positions of Pluto extracted from digitized Pulkovo photographic plates taken in 1930~--~1960.}
\author{E.V. Khrutskaya$^1$
\setcounter{footnote}{1}
\footnote{This paper is dedicated to the memory of our colleague and initiator of this work, Dr. Evgeniya Khrutskaya, who recently passed away.},
\setcounter{footnote}{0}
 J.-P. De Cuyper$^2$, S.I. Kalinin$^1$, \\A.A. Berezhnoy$^1$, G. de Decker$^2$\\\\
{\small $^1$ Pulkovo Observatory, 65/1 Pulkovskoye chaussee, Saint Petersburg, 196140, Russia\footnote{e-mail: deimos@gao.spb.ru}}\\
{\small $^2$ Royal Observatory of Belgium, Ringlaan 3, B-1180 Ukkel, Belgium}}

\begin{document}

\maketitle

\begin{abstract}
We present the results of determination of Pluto's positions derived from photographic plates taken in 1930 -- 1960. Observations were made with Normal Astrograph at Pulkovo Observatory. Digitization of these plates was performed with high precision scanner at Royal Observatory of Belgium (ROB Digitizer). Mean values of standard errors of plate positions ($x,y$) lie between 12 and 18 mas. The UCAC4 catalogue was used as an astrometric calibrator. Standard errors of equatorial coordinates obtained are within 85 to 100~mas. Final table contains 63 positions of Pluto referred to the HCRF/UCAC4 frame.
\end{abstract}

\section{Introduction}
Progress in construction of the modern planetary ephemerides systems is significantly depend on length and quality of series of positional observations of Solar system bodies. Especially this is actual for outer planets. Ephemeris based positions of Pluto traditionally demonstrate deficiencies of various dynamical ephemerides systems like DE, INPOP and other. 

Large sets of old photographic observations of Pluto were initially processed through manual measurements with various machines.  Pluto's positions observed were formally referred to various reference frames (FK3, FK4, FK5) and they were distorted by systematic errors of old reference catalogues. Hence, standard errors of these positions of Pluto were about 0.2 to 0.4~arcsec.

Significant improvement of the data considered has been made possible through digitization of old photographic plates with high precision scanners. Modern astrometric catalogues also facilitate to determine all positions of Pluto in present time reference frame (HCRF). As a result, the accuracy of the Pluto's positions refined should be about 100~mas and better. This accuracy is mainly limited due to the quality of old photographic plates and systematic errors of coordinates and proper motions of the reference stars. 

High precision positions of Pluto extracted from early photographic observations may be needed for New Horizons mission (\cite{Guo_Farquhar2005}, \cite{Beauvalet_et_al2012}).

Large archive of old photographic plates is stored at Pulkovo Observatory. These plates were mainly taken at Pulkovo Observatory. Significant part of this archive consists of the photographic plates taken to determine the positions of the Solar system bodies. Pluto's images are contained on the more than 250 photographic plates. The results of digitization and astrometric reductions of 63 early photographic plates taken at Pulkovo Observatory in 1930 -- 1960 are presented in this paper. This work is a part of large plan of digitization of old photographic plates, which is being realised at Pulkovo Observatory (\cite{Khrutskaya2013}).  A short description of astrograph and other details of observations are given in Section~\ref{obs}. Digitization of plates is schematically considered in Section~\ref{dig}. Basic stages of processing of the observations are reflected in Section~\ref{astred}. Section~\ref{finalpos} contains table of final Pluto's positions and necessary remarks. A brief overview of the main 
conclusions is given in Section~\ref{concl}.

\section{Observations}\label{obs}
The photographic observations of Pluto considered were started at Pulkovo Observatory in 1930. First plates were taken by S.G. Kostinsky in March of 1930. Observations were made every year during the Pluto's opposition with Normal Astrograph (D/F = 330~mm/3467~mm, latitude = 59.771280~deg, longitude = 30.324977 deg, altitude = 77.48 meters). The size of the photographic plate was $16 \times 16$~cm (astrometrically good FOV was $2\times 2$~degrees with scale  = 59.56~ arcsec/mm). Exposure time was 1~hour. These observations had been interrupted in 1941 due to war. The lenses of the Normal Astrograph were saved. They were installed on the reconstructed telescope in 1948. The observations of Pluto were restarted in 1949.  

\section{Digitization of plates}\label{dig}
High precision scanner of Royal Observatory of Belgium (ROB Digitizer) was used to digitize Pulkovo plates, which contain images of Pluto. This machine was designed by AERO-TECH\footnote[1]{www.aerotech.com} (USA). ROB Digitiser is equipped with Schneider Xenoplan lens and BCi4 camera ($1280\times1024$~pixels, pixel sizes are $7\times7$~mkm). Maximum plate size is 35~cm. This machine is fully automated device. More detailed information about ROB Digitizer was presented in series of papers of J.-P. De Cuyper and his colleagues (\cite{DeCuyper2004}, \cite{DeCuyper2005}, \cite{DeCuyper2006}, \cite{DeCuyper2009}, \cite{DeCuyper2012}).    

\section{Astrometric reduction}\label{astred}
\begin{figure}
   \includegraphics[width=1\columnwidth]{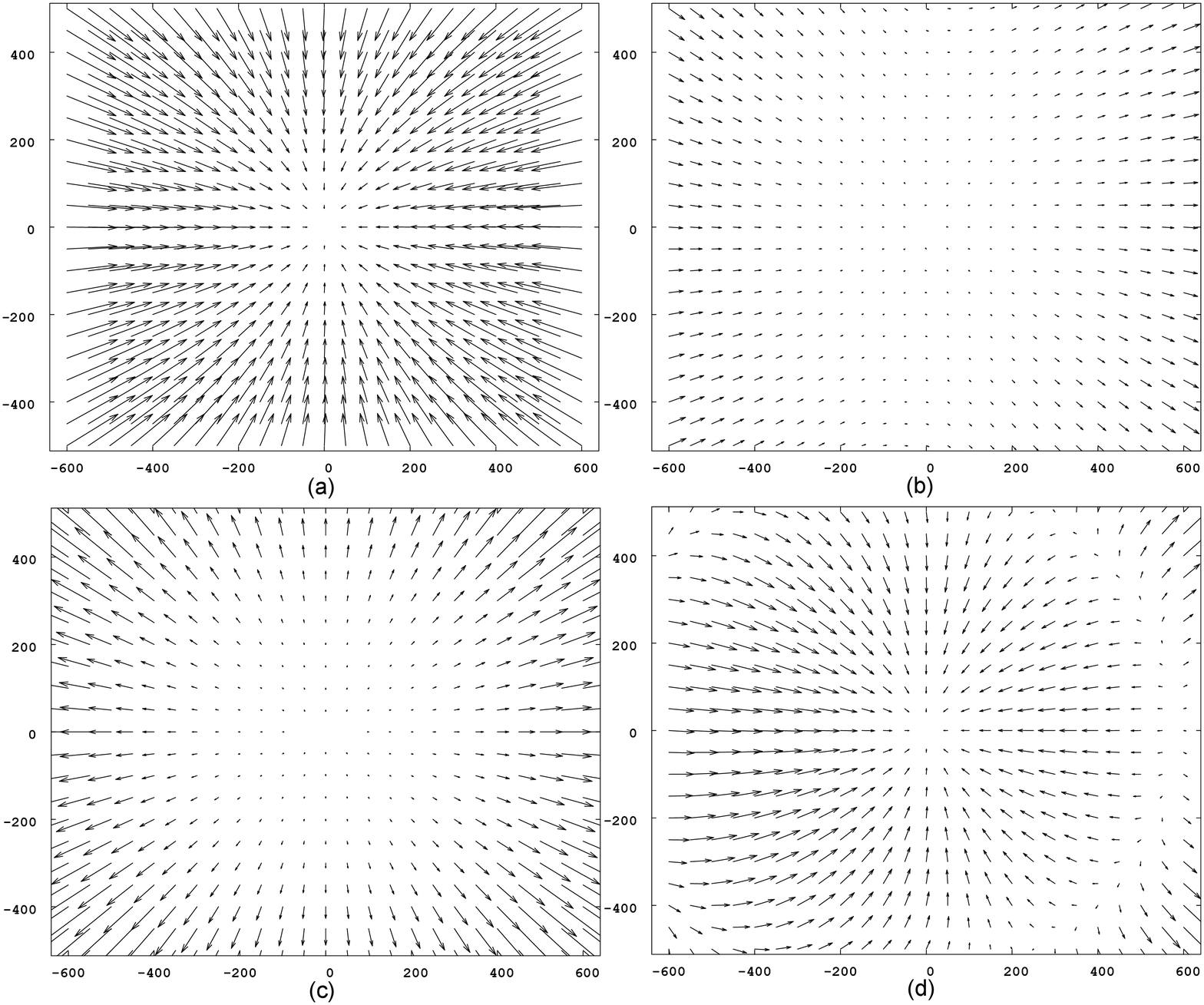}
      \caption{The components of the field of positional systematic errors caused by the camera of ROB Digitizer. The panel (a) demonstrates linear component, panel (b) shows quadratic component, and panel (c) presents the third order part. Figure (d) was constructed as a sum of
     (a),(b) and (c). Axis units are pixels. Maximum vector length is 0.25~mkm.} 
         \label{Fig2}
\end{figure}

The result of digitization of one photographic plate with the ROB Digitizer is presented as a grid of the overlapped images (imagets), which covered whole plate. The Lorentz profile was used to fit stellar and Pluto's images. Position of the centre of each imaget is determined with high accuracy (about several nm). The differences between the positions of separate star derived from two of more imagets were represented using third term polynomial model. The parameters of this model were estimated by least-squares adjustment. The components of field of positional systematic errors caused by the camera of ROB Digitizer are seen in Fig~\ref{Fig2}. As a result, plate coordinates ($x,y$) of Pluto and reference stars were corrected according to calculated parameters of adopted model of the systematic errors.     

The UCAC4 (\cite{zacharias2013}) was used as a reference catalogue. Magnitudes of the reference stars are within 11 to 14~mag. As a result, the number of reference stars was lie between 50 and 60 per one plate. Standard model of six constants was applied. Atmospheric refraction corrections were added. The unit weight errors are 90 to 100~mas. The standard errors of one Pluto's position are within 100 to 120~mas. The same values obtained using the old measuring machines and old reference catalogues were usually three or four times bigger (\cite{lavdovsky1953}, \cite{lavdovsky1968}, \cite{rylkov1996}). The attempt of refinement of the equatorial coordinates of Pluto  based on the new manual measurements of the same photographic plates was made in 1990s. New set of positions of Pluto in the HCRF/UCAC3 frame were recently published (\cite{rylkov2013}). Typical unit weight errors of astrometric reduction in this work were 230 to 330~mas. 

\section{Positions of Pluto}\label{finalpos}

The final equatorial coordinates of Pluto considered are presented in Table~\ref{Tab1}. On the whole, 63 positions have been determined in the HCRF/UCAC4 frame. The last column of Table~\ref{Tab1} contains estimations of Pluto's magnitude. These values were calculated using UCAC4 fit model magnitudes of the reference stars.  

\begin{longtable}{p{4 cm} p{3 cm} p{4 cm} p{2 cm}}
 \caption{Pulkovo positions of Pluto (1930 -- 1960).}\label{Tab1}
\\\hline
Date (UTC)& $RA_{J2000}$& $Dec_{J2000}$& mag\\
year month day& h m s& deg arcmin arcsec& \\
\hline
\endfirsthead
\caption*{Table~\ref{Tab1} -- continued}\\
\hline
Date (UTC)& $RA_{J2000}$& $Dec_{J2000}$& mag\\
year month day& h m s& deg arcmin arcsec& \\
\hline
\endhead 
1930 03 17.789719& 07 19 50.817& +21 59 35.65& 13.92\\
1930 03 30.804780& 07 19 41.127& +22 00 46.64& 15.28\\
1930 04 04.808441& 07 19 42.114& +22 01 05.24& 15.91\\
1930 04 20.828469& 07 20 02.851& +22 01 28.93& 17.00\\
1931 03 17.759216& 07 25 24.031& +22 11 58.06& 15.81\\
1932 02 26.812448& 07 31 52.265& +22 20 58.79& 15.41\\
1932 03 05.795452& 07 31 27.173& +22 22 19.08& 15.64\\
1932 03 07.812845& 07 31 21.704& +22 22 37.41& 15.22\\
1932 03 12.828627& 07 31 09.897& +22 23 18.97& 15.97\\
1933 02 21.807410& 07 37 53.523& +22 31 21.36& 15.62\\
1933 03 02.775910& 07 37 21.432& +22 33 01.61& 15.19\\
1934 03 11.821253& 07 42 45.103& +22 45 02.17& 15.48\\
1934 03 13.804711& 07 42 40.295& +22 45 17.75& 16.30\\
1935 03 25.810697& 07 48 10.850& +22 56 34.71& 15.38\\
1938 02 21.852309& 08 07 58.481& +23 17 13.83& 15.45\\
1938 02 25.801219& 08 07 41.152& +23 18 10.51& 15.66\\
1938 03 23.793942& 08 06 18.994& +23 22 23.95& 15.75\\
1939 01 19.967109& 08 17 11.937& +23 13 19.91& 14.91\\
1939 01 21.944853& 08 17 00.782& +23 14 04.09& 15.05\\
1939 02 23.865193& 08 14 07.519& +23 24 34.36& 15.21\\
1940 03 14.853653& 08 19 13.758& +23 34 43.38& 14.46\\
1940 03 25.782122& 08 18 46.056& +23 35 59.74& 15.21\\
1941 03 29.796793& 08 25 03.844& +23 41 37.95& 15.33\\
1941 04 02.810803& 08 24 57.639& +23 41 49.34& 15.45\\
1949 02 18.868166& 09 22 57.730& +23 36 16.33& 14.55\\
1949 02 26.879230& 09 22 13.618& +23 39 35.72& 14.48\\
1949 03 05.918632& 09 21 37.176& +23 42 07.97& 14.41\\
1950 02 24.875340& 09 29 47.912& +23 33 37.03& 14.98\\
1950 03 11.837299& 09 28 30.430& +23 39 04.08& 14.27\\
1950 03 14.808410& 09 28 16.579& +23 39 55.71& 14.82\\
1951 03 05.937238& 09 36 28.140& +23 30 44.55& 14.46\\
1951 03 06.957731& 09 36 22.803& +23 31 06.95& 14.60\\
1951 04 03.903170& 09 34 24.975& +23 37 42.17& 15.53\\
1951 04 07.803665& 09 34 13.827& +23 38 00.78& 14.64\\
1952 03 17.910959& 09 42 58.088& +23 27 15.01& 14.24\\
1952 04 15.875903& 09 41 21.173& +23 30 59.10& 14.65\\
1952 04 16.830595& 09 41 19.405& +23 30 57.58& 14.13\\
1952 04 19.835960& 09 41 14.359& +23 30 49.13& 14.44\\
1953 03 11.872657& 09 51 07.190& +23 16 08.47& 14.19\\
1953 03 14.872826& 09 50 52.132& +23 17 10.28& 14.61\\
1953 03 15.878425& 09 50 47.199& +23 17 29.85& 14.16\\
1954 03 21.803363& 09 58 04.735& +23 08 34.98& 14.33\\
1954 03 31.827987& 09 57 22.251& +23 10 51.52& 14.29\\
1954 04 20.844738& 09 56 24.484& +23 12 04.21& 14.54\\
1955 03 23.818454& 10 05 46.646& +22 56 56.40& 14.67\\
1955 04 08.802879& 10 04 42.566& +23 00 00.68& 14.92\\
1955 04 24.849265& 10 04 02.172& +23 00 09.34& 14.11\\
1956 03 16.883552& 10 14 13.210& +22 41 06.98& 14.14\\
1956 03 31.809652& 10 13 03.446& +22 45 24.09& 14.31\\
1956 04 30.885930& 10 11 41.737& +22 46 10.28& 14.54\\
1957 03 20.891375& 10 21 55.962& +22 26 58.64& 14.07\\
1957 03 25.864021& 10 21 31.578& +22 28 35.01& 13.76\\
1957 03 29.868455& 10 21 13.086& +22 29 40.41& 13.71\\
1957 04 01.809761& 10 21 00.274& +22 30 21.15& 14.74\\
1958 03 17.892602& 10 30 19.971& +22 08 34.30& 13.98\\
1958 03 20.875498& 10 30 04.237& +22 09 44.71& 14.15\\
1958 04 07.809527& 10 28 41.460& +22 14 37.45& 14.87\\
1959 03 03.905711& 10 39 52.973& +21 42 27.59& 14.84\\
1959 03 09.888719& 10 39 18.323& +21 45 43.26& 13.91\\
1959 03 29.844783& 10 37 31.893& +21 53 56.72& 14.07\\
1959 03 30.857340& 10 37 27.063& +21 54 14.49& 13.81\\
1960 03 21.853283& 10 46 24.607& +21 30 54.60& 15.18\\
1960 03 25.865515& 10 46 03.550& +21 32 27.32& 14.48\\

\hline
\end{longtable}

\section{Conclusion}\label{concl}
The long and homogeneous series of Pulkovo photographic observations of Pluto performed in 1930 -- 1960 and high precision scanner of Royal Observatory of Belgium have allowed us to obtain 63 accurate positions of Pluto in the HCRF/UCAC4 frame. Detailed investigation of positional systematic errors of the reference stars and Pluto have been performed. All necessary corrections were made to refine Pluto's positions. Standard errors of the final equatorial coordinates of Pluto are within 85 to 100~mas.
Only early part of Pulkovo photographic observations of Pluto has been digitized and processed. Digitization of Pulkovo photographic plates with Pluto images taken from 1960 to 1996 has been planned. Presented and expected results will be useful in construction of the modern planetary ephemerides systems.

\section*{Acknowledgements}
This work was performed with support from the Russian Foundation for Basic Research (project no. 12-02-00675a). The authors express their thanks M.Yu. Khovritchev for his help in preparing the text of the paper.

\end{document}